\documentclass[pra,superscriptaddress,twocolumn,floatfix,showpacs,preprintnumbers,lengthcheck]{revtex4-2}

\usepackage[utf8]{inputenc}
\usepackage[T1]{fontenc}
\usepackage{amsmath}
\usepackage{amssymb}
\usepackage{amsfonts}
\usepackage{newtxtext,newtxmath}
\usepackage{bm}
\usepackage{color}
\usepackage{datetime}
\usepackage{graphicx}
\usepackage{subcaption} 
\usepackage{braket}
\usepackage{float}
\usepackage{adjustbox}
\usepackage{quantikz} 
\usepackage{soul} 

\usepackage[colorlinks,citecolor=blue,urlcolor=blue,linkcolor=blue]{hyperref}

\newcommand{\ehningen}{\texttt{ibmq\_ehningen}}
\newcommand{\simulator}{\texttt{ibmq\_qasm\_simulator}}

\begin{document}
	
\title{An Energy Estimation Benchmark for Quantum Computers}
	
\date{Compiled \today, \currenttime}

\newcommand{\freiburg}{Physikalisches Institut, Albert-Ludwigs-Universit\"{a}t Freiburg, Hermann-Herder-Stra{\ss}e 3, 79104, Freiburg, Germany}
\newcommand{\eucor}{EUCOR Centre for Quantum Science and Quantum Computing, Albert-Ludwigs-Universit\"{a}t Freiburg, Hermann-Herder-Stra{\ss}e 3, 79104, Freiburg, Germany}

\author{Andreas~J.~C.~Woitzik}
\email{andreas.woitzik@physik.uni-freiburg.de}
\affiliation{\freiburg}
\author{Lukas Hoffmann}
\affiliation{\freiburg}
\author{Andreas Buchleitner}
\affiliation{\freiburg}
\affiliation{\eucor}
\author{Edoardo G.~Carnio}
\affiliation{\freiburg}
\affiliation{\eucor}

\begin{abstract}
Certifying the performance of quantum computers requires standardized tests. 
We propose a simple energy estimation benchmark that is motivated from quantum chemistry. 
With this benchmark we statistically characterize the noisy outcome of the IBM Quantum System One in Ehningen, Germany. 
We find that the benchmark results hardly correlate with the gate errors and readout errors reported for the device. 
In a time-resolved analysis, we monitor the device over several hours and find two-hour oscillations of the benchmark results, as well as outliers, which we cannot explain from the reported device status.
We then show that the implemented measurement error mitigation techniques cannot resolve these oscillations, which suggests that deviations from the theoretical benchmark outcome statistics do not stem solely from the measurement noise of the device.
\end{abstract}

\maketitle

\section{Introduction}
The physical platform where one chooses to encode information determines the rules of computation~\cite{landauer1991information}.
Based on this premise, quantum computers were proposed in the 1980s~\cite{Feynman_1982} and are intensely researched today.
The pursued applications of such devices range from the simulation of quantum systems~\cite{bharti2022noisy, cao2019quantum} to a more efficient solution of certain optimization problems~\cite{orus2019quantum}.
Several algorithms~\cite{Nielsen_2012, montanaro2016quantum}, like the one suggested by Grover~\cite{grover1996fast}, are proven to bear a speed-up on ideal (i.e., noiseless) quantum computers over classical computers.
However, the size and noise levels of current quantum hardware~\cite{bharti2022noisy} do not support the reliable execution of useful algorithms with an observable speed-up.
Currently, significant efforts are devoted to improving hardware-related~\cite{de2021materials, kjaergaard2020superconducting, blatt2012quantum, bruzewicz2019trapped}, as well as algorithmic aspects~\cite{montanaro2016quantum, bharti2022noisy, callison2022hybrid, cerezo2021variational}.
To compare the level of reliability of current error-prone quantum devices, benchmarks, i.e., predefined tasks to assess their performance, are necessary~\cite{huppler2009art}.
The latter is the reason why the interest in benchmarks for quantum hardware with few qubits and high error rates, for which error correction~\cite{Nielsen_2012} is still out of reach, has increased significantly in the past years.
Different types of benchmarks exist in the field of quantum information processing.
Early ideas of a state-preparation benchmark, gauging the fidelity of preparing a qubit register in a given target state, emerged already in the late 1990s \cite{knillCatStateBenchmarkSeven1999}.
Some benchmarks are aimed at specific applications, for example the Shor algorithm~\cite{amicoExperimentalStudyShor2019}, whereas others focus on gate errors~\cite{michielsenBenchmarkingGatebasedQuantum2017}, artificial circuits of constant width and depth~\cite{blume-kohoutVolumetricFrameworkQuantum2020}, or scalability in register size~\cite{magesanCharacterizingQuantumGates2012, cornelissenScalableBenchmarksGateBased2021}. 
Popular algorithms on current quantum computing hardware are quantum chemistry calculations \cite{cao2019quantum}, which inspire related benchmarks~\cite{mccaskeyQuantumChemistryBenchmark2019}, including the one we present and apply here to an IBM Quantum System One machine.

Our work is organized as follows: first we define our benchmark, introduce a standard measurement error mitigation technique and describe IBM Quantum System One in Ehningen, Germany in Sec.~\ref{sec:methodologyandterminology}. In Sec.~\ref{subsec:datacollection} we describe the data collection, which we analyze as a whole in Sec.~\ref{subsec:static-features}.
In Sec.~\ref{subsec:timedependentanalysis} we resolve the benchmark results in time and unveil unexpected dynamical features.  We draw conclusions about the noisy statistics of the machine in Sec.~\ref{sec:conclusion}.

\section{Methodology \& Terminology}
\label{sec:methodologyandterminology}

\subsection{$\ket{W}$-benchmark}
\label{subsec:wbenchmark}

Our benchmark revolves around the creation of a $\ket{W}$ state~\cite{dur2000three} of three qubits (i.e., two-level systems):
\begin{align}\label{eq:w-state}
\ket{W} = \frac{1}{\sqrt{3}} (\ket{001} + \ket{010} + \ket{100}) ,
\end{align}
with $ \ket{0} $ and $ \ket{1} $ the eigenstates corresponding, respectively, to the $ +1 $ and $ -1 $ eigenvalues of the Pauli $ Z $ operator. 
While this state is usually investigated for its (multipartite) entanglement properties \cite{mintert2005measures, guhne2009entanglement}, in this benchmark, it features as an eigenstate of the one-dimensional three-qubit spinless Fermi-Hubbard model with periodic boundary conditions, hopping term $t$ and vanishing on-site interaction. In second quantization, the Hamiltonian of this model reads
\begin{align}
H_{FH} = -t \sum_{\langle i,j \rangle} \left(\hat{c}_{i}^\dag \hat{c}_{j} + \hat{c}_{j}^\dag \hat{c}_i \right),
\end{align}
where the sum runs over all neighbouring sites and $\hat{c}_i, \hat{c}_i^\dag$ are, respectively, the fermionic annihilation and creation operators.
In the following, we give all energies in units of $ t $.

After a Jordan-Wigner transformation~\cite{jordan1993paulische}, the spinless Fermi-Hubbard Hamiltonian $H_{FH}$ for $n$ sites becomes
\begin{align}\label{eq:fermiHubbardHamiltonian}
H_{FH} = -\frac{1}{2} \left( Y_1 \tilde{Z} Y_n + X_1 \tilde{Z} X_n +  \sum_{j=1}^{n-1} Y_jY_{j+1} + X_jX_{j+1} \right), 
\end{align}
where $X_j, Y_j, Z_j$ are the Pauli operators (also known as $\sigma_x, \sigma_y, \sigma_z$) on the $j$-th subspace (extended, implicitly, with identity operators $ \mathbb{I}_{m\neq j} $ on all other subspaces). For convenience of notation we have set $ \tilde{Z} = \prod_{j=2}^{n-1} Z_j $.
We refer to (tensor) products of Pauli and identity operators as \emph{Pauli strings}.
Specifically for the three-qubits case which we investigate in this article, the resulting problem is the so-called \textit{Fermionic Triangle} \cite{woitzikEntanglementProductionConvergence2020}, with Hamiltonian 
\begin{equation}\label{eq:fermiTriangle}
H_{FT} = -\frac{1}{2} ( Y_1Z_2Y_3 + X_1Z_2X_3 + Y_1Y_2 + X_1X_2 + Y_2Y_3 + X_2X_3 ) .
\end{equation}

Direct calculation shows that $\ket{W}$ is the ground state of $ H_{FT} $ with eigenvalue
\begin{equation}\label{eq:energyExpectationValue}
	E = \bra{W} H_{FT} \ket{W} = -2.
\end{equation}

The benchmark which we propose consists of \emph{preparing} the qubit register, starting from the initial state $\ket{000}$, in a $ \ket{W} $ state, and then \emph{measuring} with respect to the objective function $\bra{W} H_{FT} \ket{W}$, that is its energy expectation value. In a perfectly isolated and controlled qubit register, we expect to obtain $ -2 $, as in Eq.\ \eqref{eq:energyExpectationValue}.
In a real execution on the quantum hardware, however, every gate carries an error. The influence of individual gate errors on the energy then accumulates throughout preparation and measurement. Consequently, the deviation of the objective function from the theoretical expectation will reflect the imperfections in preparation and measurement~\cite{brugger2022output}.

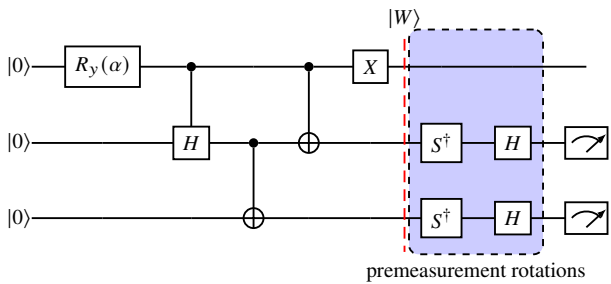
\begin{figure}
\begin{adjustbox}{width=0.46\textwidth}
\begin{quantikz}
	\ket{0} & \gate{R_y(\alpha)} & \ctrl{1} & \qw & \ctrl{1} & \gate{X} \slice{$\ket{W}$} & \qw \gategroup[3,steps=2,style={dashed,
		rounded corners,fill=blue!20, inner xsep=2pt}, background, label style={label position=below,anchor=
		north,yshift=-0.2cm}]{{premeasurement rotations}}  & \qw & \qw \\
	\ket{0} & \qw & \gate{H} & \ctrl{1} & \targ{} & \qw & \gate{S^\dag} & \gate{H} & \meter{} \\
	\ket{0} & \qw & \qw & \targ{} & \qw & \qw & \gate{S^\dag} & \gate{H} & \meter{} 
\end{quantikz}
\end{adjustbox}
	\caption{Circuit for the preparation of the $\ket{W}$ state of Eq.~\eqref{eq:w-state} (left, up to the dashed red line), and the premeasurement rotations (right, enclosed in the dashed box) needed to determine the expectation value $\bra{W} Y_2Y_3 \ket{W}$ of the Pauli string $Y_2Y_3$. The gate $R_y(\alpha)$ describes a rotation of $\alpha$ about the $y$ axis on the Bloch sphere of the relevant qubit, in our case $ \alpha = 2 \arccos(1/\sqrt{3}) $. As for the premeasurement rotations, $S$ describes a $\pi /2$ rotation about the $z$ axis of the single-qubit Bloch sphere, and $H$ is the Hadamard gate. It thus follows that $Y_j = S H Z_j H S^\dag$.}
	\label{fig:circuit}

\end{figure}

We prepare the $ \ket{W} $ state with the circuit depicted in Fig.~\ref{fig:circuit}. The preparation circuit needs to be transpiled, i.e.\ converted into a (usually longer) sequence of gates that are physically implemented on the quantum hardware at hand \cite{qiskit_transpiler}. As for the measurement, the evaluation of expectation values of Pauli strings in Eq.\ \eqref{eq:fermiTriangle} involving $ X_j $ or $ Y_j $ requires additional (premeasurement) gates implementing a rotation into the measurement basis of the device (customarily defining the eigenbasis of $ Z_j $).
In Fig.~\ref{fig:circuit}, we show the premeasurement gates needed to evaluate the expectation value $ \langle Y_2 Y_3 \rangle $.

\subsection{Measurement error mitigation}
\label{subsec:errorMitigation}

By \emph{error mitigation}, we refer to techniques developed to infer the correct result of the execution of an algorithm on real quantum hardware from its actual output data, assuming that the latter is garnished by systematic errors amenable to paradigmatic modelling.
Usually, error mitigation techniques post-process the raw output data~\cite{cai2022quantum}, in contrast to bona fide quantum error correction, where multiple physical qubits are used to encode a single logical qubit~\cite{terhal2015quantum}.
Mitigation techniques can aim for gate errors~\cite{he2020zero}, as well as for initialization or measurement errors~\cite{kwon2020hybrid}. Since the latter are dominant in small systems~\cite{nationScalableMitigationMeasurement2021}, we discuss the impact of \emph{measurement} error mitigation only.
For larger systems, gate errors might contribute equally, if not dominantly, to the accumulated errors in the system. 

For our analysis, we use the measurement error mitigation procedure implemented in \textsc{qiskit} (IBM's software development kit for its quantum devices) \cite{qiskit2021} which is based on quantum detector tomography \cite{maciejewskiMitigationReadoutNoise2020}. Here, all computational basis states $\ket{\psi_j}$ are prepared and measured on the quantum device, yielding an overlap matrix $\Lambda_{ij} = \braket{\psi_i | \psi_j}$ characterizing the measurement process (the detector). In the absence of measurement errors, $ \Lambda $ is the identity matrix, otherwise it will have off-diagonal elements. If $ \Lambda $ is invertible, its inverse can be applied to the outcome histogram $ \bold{p}_\text{exp} $ from the device to retrieve the ideal (quasi)distribution $ \bold{p}_\text{ideal} = \Lambda^{-1}\bold{p}_\text{exp} $. Inevitably, the number of sampling operations (e.g., calculation of expectation values) necessary, in this procedure, to fully determine a dense $ \Lambda $ scales exponentially with the register size, a fundamental limit of error mitigation techniques discussed, e.g., in Ref.~\cite{takagiFundamentalLimitsQuantum2022}.

\subsection{Platforms}
\label{subsec:ibmqehningen}

In what follows we will use the string \ehningen{} to indicate the IBM Quantum System One machine located in Ehningen, Germany. This device is part of IBM's Falcon family of machines with 27 qubits and sports variant r5.11 \cite{falcon}.
Communication with the device is possible via a web interface, and computations run in deferred time managed by a scheduler.

The string \simulator{}, instead, will refer to a fictitious device (provided by IBM's qiskit) that simulates quantum computations in circuit form, but with perfect readout and gate operations.

\section{Numerical Results}
\label{sec:numericalresults}

\subsection{Data collection}
\label{subsec:datacollection}
Let us now define the vocabulary describing the data collection process.
First of all, one \emph{realization} of the objective function in Eq.~\eqref{eq:energyExpectationValue} is determined by summing the expectation value of each individual Pauli string in the Hamiltonian (i.e., of the six strings in Eq.~\eqref{eq:fermiTriangle}). Each expectation value is determined in an \textit{experiment}. An experiment consists of 1024 \textit{shots}, where a shot consists of the initialization of the initial three-qubit state $\ket{000}$, the execution, on the chosen platform (simulator or quantum hardware), of the transpiled version of the unitary depicted in Fig.~\ref{fig:circuit}, the premeasurement rotations and the measurements in the measurement basis of the device. Each shot yields a three-bit string describing the measurement outcome of $ Z_1 Z_2 Z_3 $ on the prepared and rotated (during the premeasurement stage, see Fig.~\ref{fig:circuit}) state produced by the circuit. An experiment therefore collects a histogram of the outcome statistics and uses it to compute the expectation value of the given Pauli string, as prescribed by a quantum mechanical state model. The thus obtained expectation values for the Pauli strings entering~\eqref{eq:fermiTriangle} are then added to form the energy expectation value~\eqref{eq:energyExpectationValue} that is determined by one realization.

In each \textit{job} we evaluated repeatedly for several hours the objective function $\bra{W}H_{FH}\ket{W}$ on one or more qubit triplets, i.e., randomly chosen subsets of three connected qubits (notice that the same triplet might be probed in different jobs). The time-dependent data on one triplet from one job we call a \textit{time series}. We store \textit{packets} of 50 realizations, the corresponding expectation values of the individual Pauli strings, as well as the timestamp and the device properties upon generation of a new packet of results, as provided by the machine interface. Further properties of interest for us will be the readout and gate errors, which are updated only at calibration~\cite{ibm_calibration}. Since no calibration can take place during a job, these figures are reported constant within each time series. The average error rates $\epsilon = 1 - F$, where $F$ is the average gate fidelity~\cite{emerson2005scalable, magesan2012characterizing}, reported by the device interface are on the order of $10^{-4}$ for one-qubit gates, $10^{-2}$ for two-qubit gates, and $10^{-2}$ for measurements.

All data was collected in February, March and September to December of 2022, for a total of $124950$ realizations in 72 time series. In the following we comment first on the distribution of all realizations, and then on their fluctuations in time.

\subsection{Statistics of the whole data set}\label{subsec:static-features}

\begin{figure}
  \includegraphics[width=\columnwidth]{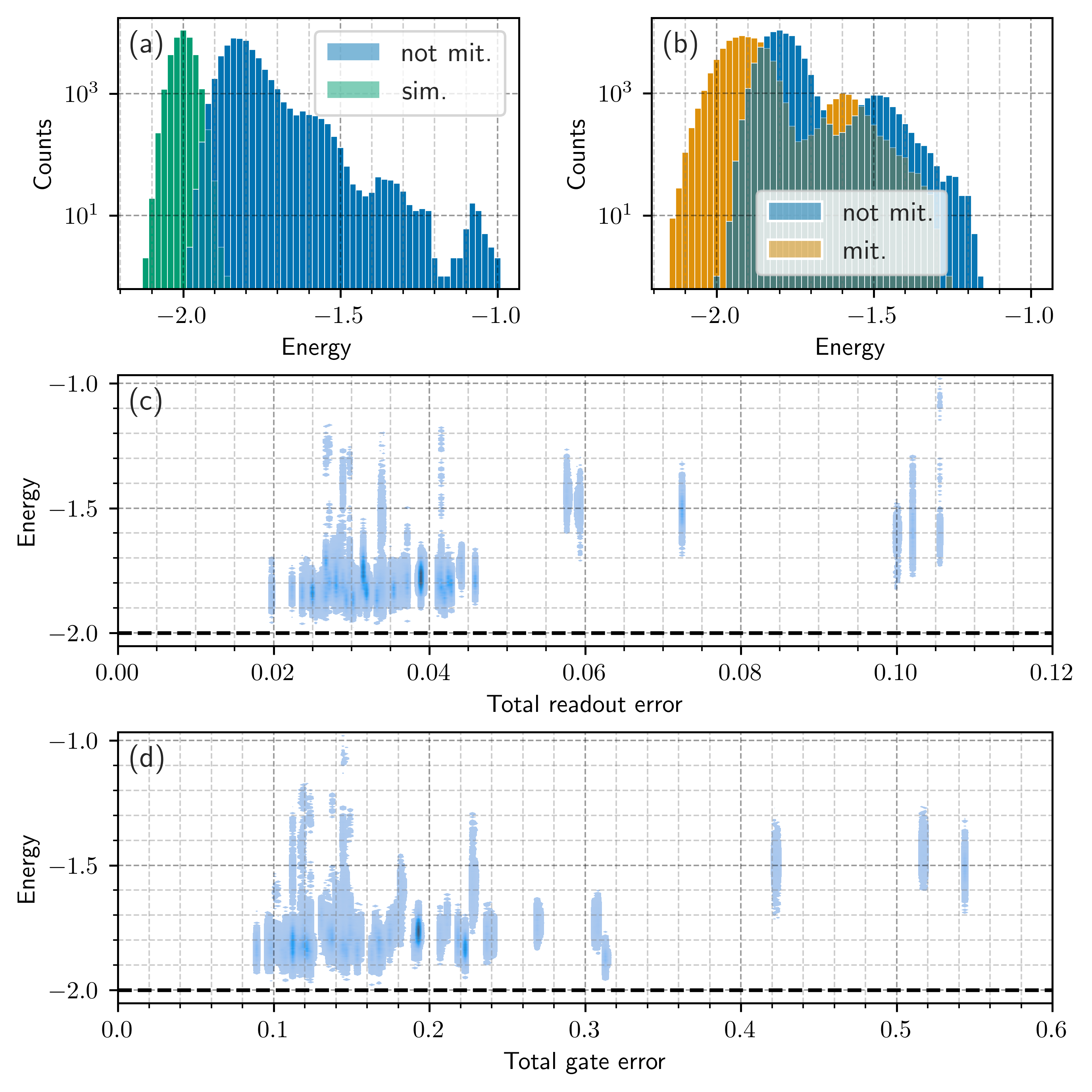}
	\caption{(a) Comparison between the histogram of the unmitigated energies sampled on \texttt{ibmq\_ehningen} in February and March 2022 (blue, ``not mit.'') and of the 39750 energies sampled on \texttt{ibmq\_qasm\_simulator} (green, ``sim.'').
	(b) Comparison between the histograms of the energies sampled on \texttt{ibmq\_ehningen} between September and December 2022 (blue, ``not mit.'') and the correspondingly mitigated values (orange, ``mit.'').
	(c,d) Density plots of the sampled energies measured on \ehningen{}, discriminated on the abscissae by (c) the total readout error and (d) the total gate error, calculated, for each job, by summing over the corresponding errors reported by the device interface at the time of calculation. The dashed line at $ -2 $ indicates the expected value from Eq.~\eqref{eq:energyExpectationValue}. Darker (lighter) colours indicate a higher (lower) concentration of samples where a given energy with associated total gate or readout error was recorded. Adapted from~\cite{hoffmann2022}.
	}
	\label{fig:allDataHistogram}
\end{figure}

In Fig.~\ref{fig:allDataHistogram}(a), we compare the histograms of realizations collected on \ehningen{} (blue) and on \simulator{} (green).
The \simulator{} histogram is highly symmetric and peaks at $ -2.00 $ with a standard deviation of $ 0.03 $. 
Since this platform simulates error-free gates, the width of this histogram descends from the inevitable rounding that occurs when estimating the expectation value of $ 2/3 $ of each Pauli string with a finite number of shots.
These small deviations are further propagated by summing over the expectation values to obtain the expected energy, which can then result lower than $ -2 $ without truly violating the variational principle. 

The \ehningen{} histogram peaks at around $-1.83$, and is significantly skewed, with a tail extending to values as large as $ -1 $.
The deviation from \simulator{} can be again traced back to the statistics of the individual Pauli strings.
As Table~\ref{table:outcome-statistics} shows for a qubit triplet selected from an exemplary job, the occurrence rates of bit strings that are summed with a positive (negative) sign are - with only two exceptions - consistently smaller (larger) than the exact rate. These systematic shifts imply that, in total, the expectation value of each Pauli string never reaches $ 2/3 $, and, hence, that, compared to \simulator{}, the distribution of the sampled energies is shifted to higher values (due to the global minus sign of the Hamiltonian~\eqref{eq:fermiTriangle}). Moreover, the sampled energies tend towards $ 0 $,  when all bit strings occur with the same frequency, i.e., when the register is prepared in (or decoheres to) the maximally mixed state.

Let us now try to resolve the impact of gate and of readout errors separately. To model a readout error (a `0' is misread as `1', and vice versa), we consider the probability $ p $ of a bit flip $ X $ on any of the $n$ qubits~\cite{Nielsen_2012}:
\begin{equation}\label{eq:bit-flip}
	\Lambda = \bigotimes_{j=1}^n \left[(1-p)\mathbb{I}_j + p X_j \right] = (1-np) \mathbb{I} +p \sum_{j=1}^n X_j + \mathcal{O}(p^2),
\end{equation}
where we have used $ (1-p)^n \approx (1-n p) $, as long as ${ p, np \ll 1 }$, and the identity $ \mathbb{I} $ acts on all qubits. In this elementary model we can estimate $ 1-np $ by averaging over the diagonal elements of the matrix $ \Lambda $ (cf.~\ref{subsec:errorMitigation}), estimated right before and for each packet of 50 realizations. For the statistics presented in Table \ref{table:outcome-statistics}, we obtained $ {p \approx 1.1 \%} $, which is in line with the reported error rate.
In Fig.~\ref{fig:allDataHistogram}(c), we see that this estimate is compatible with the distribution of the \emph{total} readout errors, i.e., the sum of the readout error of each qubit, comparable to $ np $.
If we interpret $ np $ as the error, in the sense of a standard deviation, on the expectation values of the Pauli string, and treat these as independent and identically distributed random variables, we obtain a propagated readout error of $ 3 \times 1.1 \% \times \sqrt{6} \approx 8 \% $, compatible with the shift in the histograms shown in Fig.~\ref{fig:allDataHistogram}(b) achieved via measurement error mitigation. Given the statistical assumptions above, as well as the hypotheses underpinning any error-mitigation method, this figure must be taken as a rough estimate.

The total gate error, shown in Fig.~\ref{fig:allDataHistogram}(d), is the sum of the errors of all gates involved in a realization, i.e., from the preparation and (pre)measurement of each experiment, see Fig.~\ref{fig:circuit}. As a rule of thumb, this figure is dominated by the error of the controlled (two-qubit) gates, which appear in the preparation stage of the circuit. We assume, then, that the total gate error determines the width of the histograms in Fig.\ref{fig:allDataHistogram}(a,b). This conclusion is inspired by the observation that mitigation of readout errors in Fig.~\ref{fig:allDataHistogram}(b) shifts the histogram significantly, but barely affects its width or overall shape, especially in the tail to higher values.

Based on Fig.~\ref{fig:allDataHistogram}(c,d), a clear quantitative relation between total readout or gate error and benchmark result is hard to establish. It is important to note, however, that, by interface design, the device returns constant values of these quantifiers throughout a job of several hours. In other words, the state of the device is formally considered \emph{static} between calibrations. In the next section we therefore test this assumption by analyzing whether and how the value of the objective function changes in time.

\begin{table*}
	\begin{tabular}{c||cccc|c||cc|c||c}
		Bit string\textbackslash Pauli string & $Y_1 Y_2$ & $X_1X_2$ & $Y_2Y_3$ & $X_2X_3$ & \textit{Exact} & $Y_1Z_2Y_3$ & $X_1Z_2X_3$ & \textit{Exact} & $E$ \\
		\hline
		$+000$ & 0.41(2) & 0.41(2) & 0.41(2) & 0.41(2) & 5/12 & 0.32(1) & 0.31(1) & 1/3\\
		$-001$ &  &  & 0.091(9) & 0.099(9) & 1/12 & 0.022(5) & 0.019(5) & 0\\
		$-010$ & 0.095(9) & 0.095(9) & 0.093(9) & 0.10(1) & 1/12 & 0.088(9) & 0.095(9) & 1/12\\
		$+011$ &  &  & 0.40(2) & 0.39(2) & 5/12 & 0.080(9) & 0.081(8) & 1/12\\
		$-100$ & 0.089(9) & 0.10(1) &  &  & 1/12 & 0.015(4) & 0.015(4) & 0\\
		$+101$ &  &  &  &  & 0 & 0.31(1) & 0.32(1) & 1/3\\
		$+110$ & 0.40(2) & 0.39(1) &  &  & 5/12 & 0.079(8) & 0.091(9) & 1/12\\
		$-111$ &  &  &  &  & 0 & 0.087(9) & 0.066(8) & 1/12\\
		\hline
		$\langle \cdots \rangle$ & 0.63(3) & 0.60(3) & 0.63(3) & 0.60(3) & 2/3 & 0.58(3) & 0.61(3) & 2/3 & $-1.83(3)$ \\
	\end{tabular}
\caption{Outcome statistics for qubits 11, 14 and 13 of \texttt{ibmq\_ehningen} (job started on 2022-09-19 at 22:47:16 with a duration of 7.5 hours). The data contains 31 packets of 50 sampled energies $E = \langle H_{FT} \rangle$ from Eq.~\eqref{eq:fermiTriangle}. The first column lists the possible outcomes of the circuit in Fig.~\ref{fig:circuit}, with the sign (positive or negative for, respectively, even or odd number of `1' in the bit string) indicating how the contributions need to be summed for the expectation value $ \langle \cdots \rangle $ in the last row. The central columns report the occurrence frequencies (with their standard deviation in brackets) of each possible measurement outcome for each of the Pauli strings listed in the top row. To the right of the Pauli strings involving 2 and 3 qubits, respectively, the \emph{exact} probabilities predicted by quantum mechanics are given. In the bottom right corner of the table, we compute the expectation value of the energy. Values and uncertainties are, respectively, averages and standard deviations we extracted from the histograms collected for each possible measurement outcome, and for each Pauli string.}
\label{table:outcome-statistics}
\end{table*}

\subsection{Time-resolved features}
\label{subsec:timedependentanalysis}

\begin{figure*}
	\includegraphics[width=\textwidth]{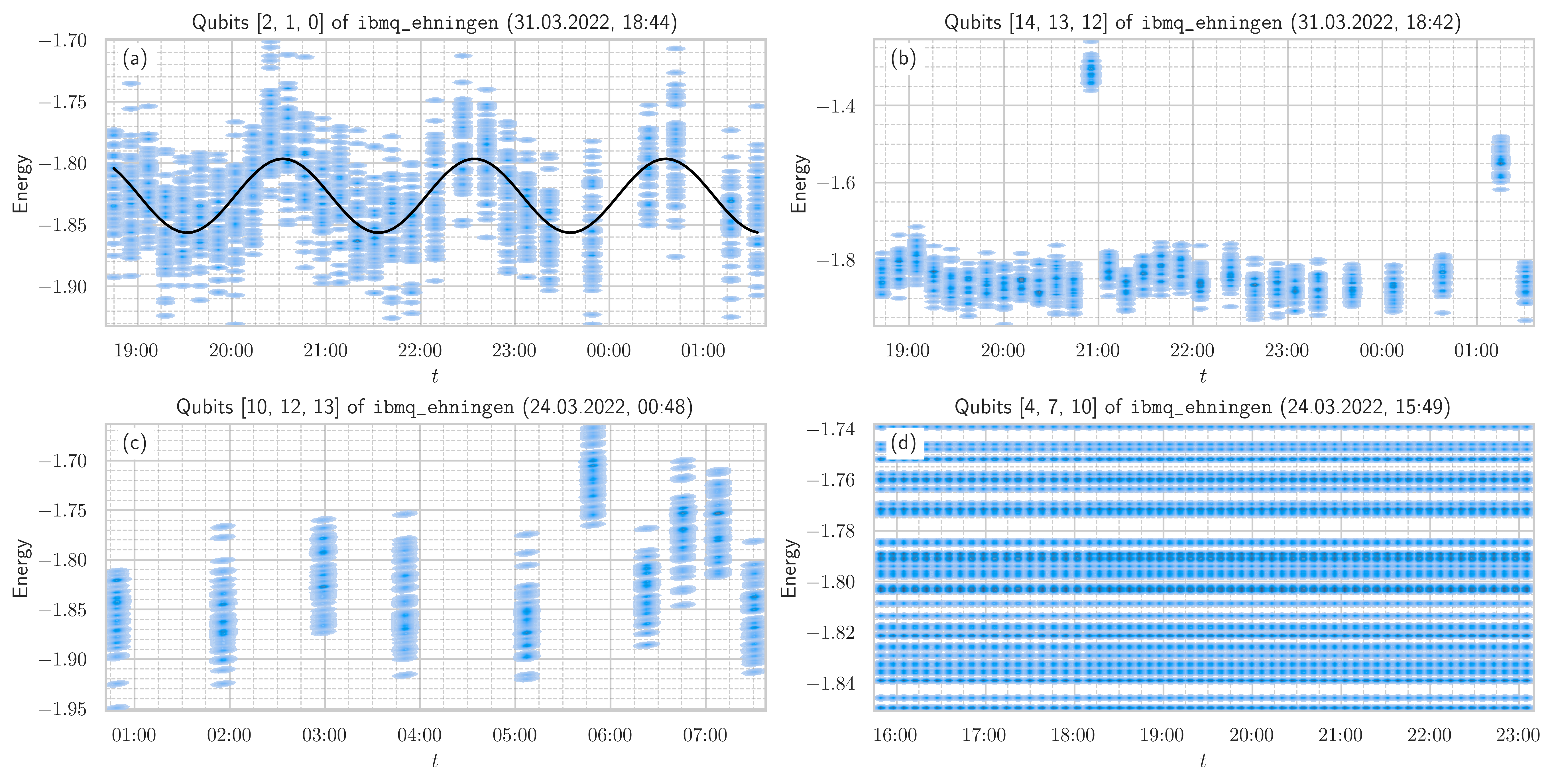}
	\caption{Time dependence of the sampled ground state energies of~\eqref{eq:energyExpectationValue} exhibiting different phenomena:
	(a) oscillations with a period of roughly two hours [$121.8(9)$ minutes in the example shown], (b) outliers, (c) delays and (d) constant results. The black line in (a) guides the eye with a sinusoidal fit of the data. All data shown is not mitigated. Adapted from~\cite{hoffmann2022}.}
\label{fig:phenomenology}
\end{figure*}

As discussed in the previous section, the outcome of each shot is intrinsically stochastic. For this reason, the baseline behaviour that we expect to, and indeed mostly do, observe are fluctuations compatible with the histograms in Fig.~\ref{fig:allDataHistogram}(a,b). However, resolving the data on the time axis also reveals unexpected phenomena that we would not be able to observe in the collective analysis above. In Fig.~\ref{fig:phenomenology} we plot the density histograms in energy (ordinates) as a function of the recorded timestamps (abscissae). We can discern:
\begin{enumerate}
\item \textit{Oscillations.} Periodic oscillations of the average energy, with amplitude smaller than the typical width of the density histograms, can be qualitatively noted for the majority of time series (see, e.g., Fig.~\ref{fig:phenomenology}(b)). However, a significant portion (22 out of 72 analyzed time series) shows oscillations with a far larger amplitude. An example is shown in Fig.~\ref{fig:phenomenology}(a), where we fit, using SciPy's~\cite{2020SciPy-NMeth} least-squared method, a model of the form $y_0 + A \sin[t/(2\pi T)+\phi_0]$, and extract a period of $T \approx 2\,\text{h}$. These oscillations were observed both in Spring and in Autumn 2022, and independently of qubit allocation. A qualitative estimation finds a period of 1 to $2\,\text{h}$ for all affected time series. In few cases the signal presented an additional positive or negative slope, suggesting the presence of oscillations on even longer time scales.
\item \textit{Outliers.} In several cases (21 out of 72) we observe time series where packets of 50 samples produce atypically high values (compared to the long-time bulk behaviour) of the objective function. An example is shown in Fig.~\ref{fig:phenomenology}(b). In particular, we observed this behaviour in jobs where we collected data from distinct qubit triplets one after the other (in a rotating scheme): outliers appeared in the same time bin for triplets with one or more qubits in common, but not for other (uncorrelated) triplets. This observation suggests that, during the job, some event affected one or more qubits, but not necessarily the whole device. Curiously, such an event involves whole packets whose density histogram is shifted to higher energy values, but is not broadened.
\item \textit{Delays.} A few time series, affecting in total 33 out of 72 tested registers, showed inconsistent time resolution, with, e.g., sparser sampling at the beginning of the time series [Fig.~\ref{fig:phenomenology}(c)], or at the end  [Fig.~\ref{fig:phenomenology}(b)]. We cannot establish whether this phenomenon is due to a slow response of the quantum device, of the hardware around it, or to yet another reason. As mentioned above, we assume no calibration to take place during any of our jobs~\cite{ibm_calibration}.
\item \textit{Constant results.} A few jobs over several days in Spring 2022 resulted in perfectly constant results for every qubit triplet tested. An example is shown in Fig.~\ref{fig:phenomenology}(d). This behaviour most certainly resulted from a software error on IBM's side that has not occurred since. Nevertheless, since the device did not report any anomaly, this error would not have been noticed had we just collected the statistics in a histogram. The data from these jobs has been excluded from any statistical analysis.
\end{enumerate}

These time-resolved features are incompatible with the nominal information provided by the interface to the quantum device. As discussed above, most outliers seem to be due to events affecting (at least) one qubit; if these events were certified as external accidents \cite{Martinis_2021}, this might be sufficient reason to exclude outliers from the statistical analysis. 

\subsection{Mitigation of measurement errors}

\begin{figure}
	\includegraphics[width=\columnwidth]{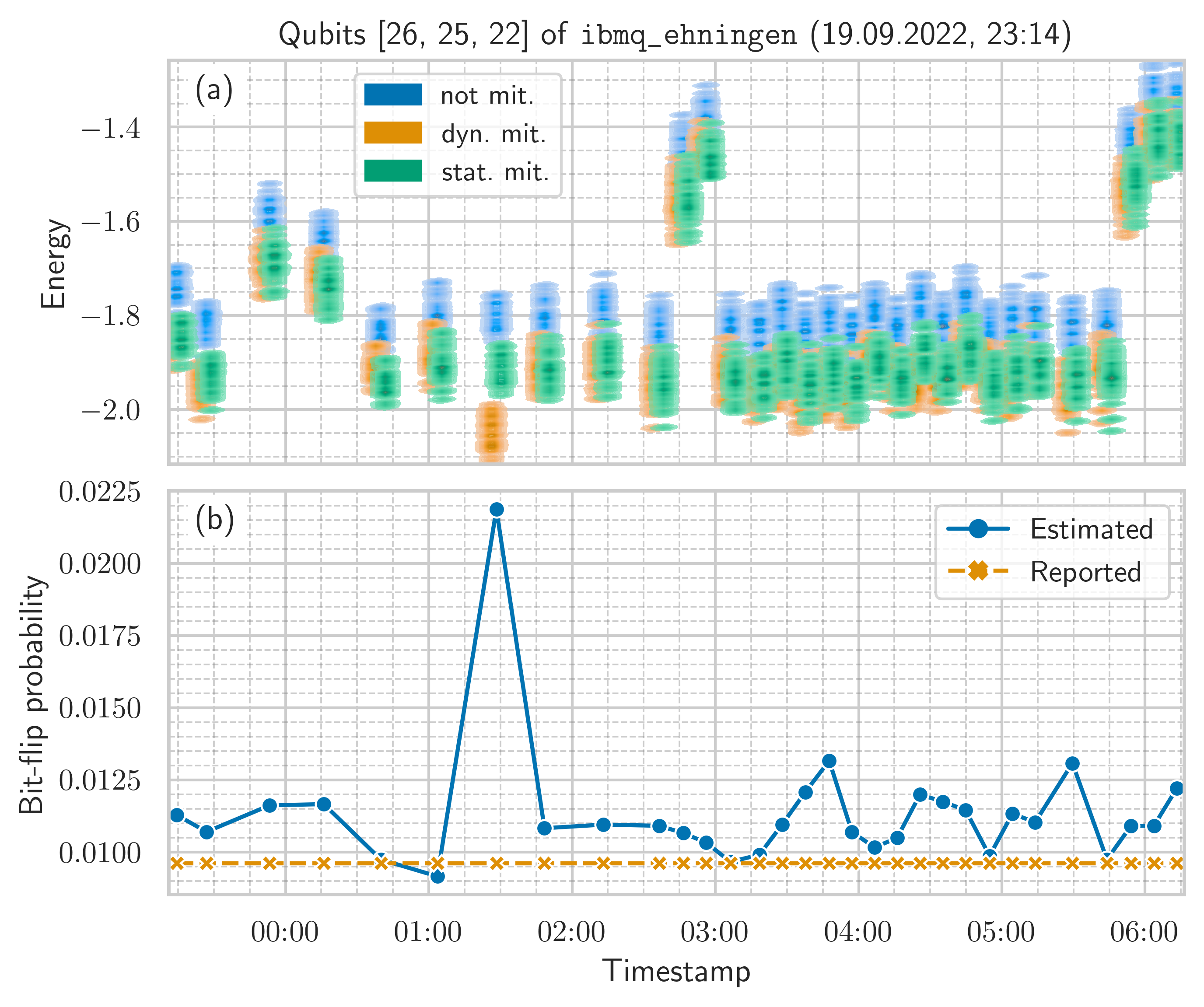}
	\caption{(a) Density histograms of the energies sampled as packets over time, see the main text for details of the data acquisition. Raw data is in blue, statically mitigated data (calibration matrix calculated once per job) in green (slightly shifted to the right for visibility), dynamically mitigated (calibration matrix calculated with each packet) in orange  (slightly shifted to the left).
	(b) In blue the bit-flip probability $ p $ estimated from the mitigation matrix, as in Eq.~\eqref{eq:bit-flip}, and in orange the average bit flip error per qubit reported by the machine.}
	\label{fig:mitigation}
\end{figure}

To discern whether the dynamics observed in Sec~\ref{sec:numericalresults}. C is caused by readout errors, we compare (measurement-error-)mitigated and unmitigated data.

When we compute the matrix $ \Lambda $ (cf.~\ref{subsec:errorMitigation}) only once, at the beginning of the job, and post-process the entire time series with it, we observe (Fig.~\ref{fig:mitigation}) a systematic shift of the sampled energies to lower values, compatible with the shift in the histograms of Fig.~\ref{fig:allDataHistogram}(b).
While this outcome is expected, it assumes static device properties throughout each job, an assumption that clashes with the observed periodic oscillations and outliers, which predictably persist despite the mitigation effort. 

To mitigate time-dependent errors, we therefore characterize the device, by means of the $ \Lambda $ matrix, before each packet of 50 samples. This ``dynamic'' mitigation is an admittedly expensive procedure, but is feasible for our three-qubit test case. 
Comparison with the ``static'' mitigation above, however, shows very little difference (see Fig.~\ref{fig:mitigation}): both, oscillations and outliers, are unaffected, although the former might be less recognizable due to the expected broadening of the mitigated energy histograms \cite{temmeErrorMitigationShortDepth2017}.

Since, via the $ \Lambda $ matrix, we have a snapshot of the device for every packet of 50 energy samples, we plot in Fig.~\ref{fig:mitigation}(b) the time-dependent estimate of the bit-flip probability $ p $ for the job presented in Fig.~\ref{fig:mitigation}(a). As discussed in Sec.~\ref{subsec:static-features}, this quantity is a proxy for the average readout measurement error per qubit. Several features are visible: (i) the average readout error reported by the device is consistently slightly lower than the estimated $ p $, (ii) after time stamp 03:00, $ p $ also exhibits oscillations, albeit on a shorter (as compared to Fig.~\ref{fig:phenomenology}(a)) timescale of approximately $1\,\textrm{h}$, (iii) there is an unexpected spike at around 01:30, which leads to unusually large mitigation of the corresponding histogram, and (iv) no features in the time-dependence of $ p $ suggest the identification of the outliers at around 03:00 and 06:00, which are therefore mitigated comparably to the other time bins.

To conclude, both static and dynamic mitigation of readout errors induce a global shift of around 5\%-8\% of the sampled energies, but do not correct for the time-dependent features observed in Fig.~\ref{fig:phenomenology}.

\section{Conclusions}
\label{sec:conclusion}

In this article, we proposed a simple energy estimation benchmark inspired from quantum chemistry and applied it to the IBM Quantum System One in Ehningen, Germany. From the benchmark outcomes we have made the following observations:
\begin{enumerate}
	\item The sampled energies (i.e., the benchmark outcomes) show a shifted and broader histogram compared to the expected values affected by quantum-mechanical noise only [Fig.~\ref{fig:allDataHistogram}(a)]; this trend, however, seems to correlate weakly with the reported total readout [Fig.~\ref{fig:allDataHistogram}(c)] or gate errors [Fig.~\ref{fig:allDataHistogram}(d)].
	\item Even though the properties of the device are nominally constant throughout the jobs, the benchmark outcomes show dynamical features beyond typical stochastic fluctuations, such as oscillations with a period of roughly $2\,\text{h}$ [Fig.~\ref{fig:phenomenology}(a)] and outliers [Fig.~\ref{fig:phenomenology}(b)].
	\item Measurement error mitigation reduces the deviation between sampled and expected energy, but it does not reduce the spread [Fig.~\ref{fig:allDataHistogram}(b)] or the time-dependence [Fig.~\ref{fig:mitigation}(a)] of the histograms.
	\item Dynamical mitigation, i.e., based on a $ \Lambda $ matrix estimated for each data sampling packet individually, does not correct for the observed phenomenology [Fig.~\ref{fig:mitigation}(a)], which therefore cannot originate from measurement errors only. In fact, the average readout error rate itself shows unexpected time-dependent features that seem uncorrelated with the dynamics observed in Fig.~\ref{fig:phenomenology}.
\end{enumerate}

From observations 3 and 4 alone it follows that the current paradigm, assuming static machine properties between calibrations, is insufficient to determine a realistic uncertainty on the machine's outcomes. 
The dynamics of error rates, as well as their accumulation across circuits and experiments, needs to be taken into careful consideration. In particular, unexpected time dependencies need to be resolved by collecting results at several time instances. 
The collection of suitable statistics, following the praxis of experimental physics, is not yet well established when using (noisy) quantum devices, certainly because of the high costs they enforce.
However, if a noisy or disordered system is the physical platform where information is encoded, ensemble statistics, recorded over reasonable time series is required to make quantitative statements of the controllable reliability.

\begin{acknowledgements}
	We thank Daniel Urban, Andreas Ketterer and Thomas Wellens for their insights on IBM Quantum System One.
	
	We thank Dominik Lentrodt for helpful feedback on the manuscript.
	
	A.~W. acknowledges the Konrad-Adenauer-Foundation for financial support.
	E.~G.~C. acknowledges support from the Georg H.~Endress foundation and from the project ``SiQuRe'' (Kompetenzzentrum Quantencomputing Baden-W\"urttemberg).
	
	Funded by the Ministerium für Wirtschaft, Arbeit und Tourismus of the State of Baden-Württemberg.
\end{acknowledgements}



%

\end{document}